\documentclass[3p,times,procedia]{elsarticle}
\usepackage{nupha_ecrc}
\volume{00}
\firstpage{1}
\journalname{Nuclear Physics A}
\runauth{Y.~Tachibana for the JETSCAPE Collaboration}
\jid{nupha}
\jnltitlelogo{Nuclear Physics A}
\usepackage{amssymb}
\usepackage[figuresright]{rotating}
\begin{document}
\begin{frontmatter}

\dochead{XXVIIIth International Conference on Ultrarelativistic Nucleus-Nucleus Collisions\\ (Quark Matter 2019)}

\title{Hydrodynamic response to jets with a source based on causal diffusion}

\author{Yasuki~Tachibana~for~the~JETSCAPE~Collaboration}

\address{Department of Physics and Astronomy, Wayne State University Detroit, Michigan 48201, USA}

\begin{abstract}
We study the medium response to jet evolution in the quark-gluon plasma within the JETSCAPE framework. Recoil partons’ medium response in the weakly coupled description is implemented in the multi-stage jet energy-loss model in the framework. As a further extension, the hydrodynamic description is rearranged to include in-medium jet transport based on a strong-coupling picture. To interface hydrodynamics with jet energy-loss models, the hydrodynamic source term is modeled by a causal formulation employing the relativistic diffusion equation. 
The jet shape and fragmentation function are studied via realistic simulations with weakly coupled recoils. We also demonstrate modifications in the medium caused by the hydrodynamic response.
\end{abstract}


\end{frontmatter}

\section{Introduction}
\label{intro}
In heavy-ion collisions, jet substructures are significantly modified via collisional and radiative processes by interactions with the Quark-Gluon Plasma (QGP). 
At the same time, medium response, the energy and momentum transportation via medium constituents excited by the jet propagation, affects the jet modification. 
Since the medium response is correlated with jets, it cannot be separated experimentally from the hard partons and should therefore be considered part of the jet.

JETSCAPE \cite{Cao:2017zih, Putschke:2019yrg, Kumar:2019bvr} is a public event generator framework for heavy-ion collisions 
which incorporates two different models describing the response of the medium. 
In the weak-coupling approach, one considers recoils in partonic scatterings with medium constituents and simulates their in-medium evolution. 
The recoils continue to scatter off other partons in the medium resulting in ever growing wakes. 
The second is a hydrodynamic prescription, applied to the soft component of wakes induced by the jet propagation using a strongly-coupled description. 
In this option, soft partons with energy below a predetermined threshold are removed from the jet shower and their energy and momentum are assumed to be locally distributed as a delta function in space-time. 
Then, a relativistic diffusion equation is employed to model the evolution of the locally distributed energy and momentum from the soft partons. 
The eventual profile of the diffused energy and momentum is used to construct a hydrodynamic source term incorporated with $(3+1)$-D viscous hydrodynamics. 

In these proceedings, we study the medium response to jet propagation and its effect via simulations within JETSCAPE. 
We present results for jet substructure observables with the effect of the medium response modeled by weakly-coupled recoils. 
In addition, we explore how the contribution from the hydrodynamic medium response can manifest itself in the final state in heavy ion collisions. 
Complementary studies by JETSCAPE including the recoil medium response for $R_\mathrm{AA}$ and $v_2$ of inclusive jets and charged hadrons are presented in Ref.~\cite{Amit:QM19}. 

\section{Medium Response in JETSCAPE}
\label{model}
\subsection{Weakly Coupled Description: Recoils}
\label{model:weak}
In the current version of JETSCAPE, the MATTER module~\cite{Majumder:2009zu,Majumder:2013re} for medium modified splittings of highly virtual partons at early stages and the LBT module~\cite{Luo:2018pto} for the later stage have a feature for taking into account the medium response effect via recoil partons. 
In this recoil prescription, all energy-momentum transfer between jets and the medium is described by individual 2-to-2 partonic scatterings, assuming weak coupling. 
Partons from the thermal medium are sampled in the scatterings with hard jet partons. 
After a scattering, the recoil parton from the medium propagates and interacts with the medium in the same way as other hard partons.
On the other hand, the recoil partons leave energy-momentum deficits (holes) in the medium. 
To ensure energy-momentum conservation, the energy and momentum of the holes are subtracted directly from the jet in the final state. 

\subsection{Strongly Coupled Description: Hydrodynamic Response}
\label{model:strong} 
In this case, we assume that the soft component of the jet, evolves hydrodynamically with the bulk medium.
This hydrodynamic medium response can be simulated by solving the hydrodynamic equation with a source term~\cite{Tachibana:2017syd,Chen:2017zte,Chang:2019sae,Tachibana:2020mtb}, 
\begin{eqnarray}
\label{eom}
\nabla_\mu T^{\mu \nu}_{\mathrm{med}}(x)=J^\nu_{\rm jet}(x), 
\end{eqnarray}
where $T^{\mu \nu}_{\mathrm{med}}$ is the energy-momentum tensor of the medium fluid, 
and $J^{\nu}_{\mathrm{jet}}$ is the source term that simulates the transfer of jet energy and momentum to the fluid. 
The contribution of hadrons originating from the hydrodynamic source term is obtained through the Cooper-Frye formula~\cite{Cooper:1974mv} along with the medium contribution and added to the final jet after the appropriate background subtraction. 

In the current version of JETSCAPE, we have introduced a module which generates the source term in Eq.~(\ref{eom}) associated with the jet energy loss models. 
When turning on the module, the holes left by recoils and soft partons with energy below a cut $E_{\mathrm{cut}}$ at the fluid's local rest frame are thermalized, and then their energy and momentum are transferred to the medium fluid via the source term. 
To construct the source term with a finite profile, we simulate the space-time evolution of the localized energy and momentum of the particles being thermalized before the transfer. 
Consider a slew of particles deposited by jets over time in the medium, enumerated by an index $i$, where the energy-momentum density emanating from each of these depositions $j_i^\nu(x)$ obeys a causal diffusion equation~\cite{Tachibana:2020mtb}:
\begin{eqnarray}
\label{diff}
\left[\frac{\partial}{\partial t}+\tau_{\rm diff}\frac{\partial^2}{\partial t^2}-D_{\rm diff}\nabla^2\right]j_{i}^\nu(x)=0. 
\end{eqnarray}
The initial condition of $j_{i}^\nu$ at the particle's creation time $t_{i}$ is given as $j_i^\nu (t = t_i,{\bf x})= p_i^\nu \delta^{(3)}({\bf x} - {\bf x}_i)$, where $p_i^\nu$ is the four momentum (holes with $p^\mu$ contribute negatively with $-p^\mu$ to the density) and ${\bf x}_i$ is the position of the particle $i$. 
To make the solution unique, another initial condition $\partial j_i^\nu/\partial t (t = t_i,{\bf x}) = 0$ is also imposed. 
The constants $D_\mathrm{diff}$ and $\tau_\mathrm{diff}$ in Eq.~(\ref{diff}) are the diffusion coefficient and the relaxation time, respectively. 
Their values are chosen to make the signal velocity $c_\mathrm{diff} = (D_\mathrm{diff}/\tau_\mathrm{diff})^{1/2}$ preserve causality, $c_\mathrm{diff}\leq 1$.

In our model, we evolve the energy-momentum density of each particle $j_i^\nu$ until a thermalization time $t_\mathrm{th}$. 
The source term is constructed with the solutions of Eq.~(\ref{diff}) as $J_{\rm jet}^{\nu}(x)=\sum_i j_i^{\nu}(x) \delta (t-[t_i + t_\mathrm{th}])$, where the sum is taken over all soft partons and holes. 
This formulation provides the source term with a finite space-time profile that naturally preserves causality. 

\section{Simulations and Results}
\label{results}
\subsection{Simulations with the Weakly Coupled Description}
\label{re:weak}
\begin{figure}[htbp]
\vspace{-14pt}
\begin{center}
\includegraphics[width=0.998\textwidth,bb=0 0 1528 625]{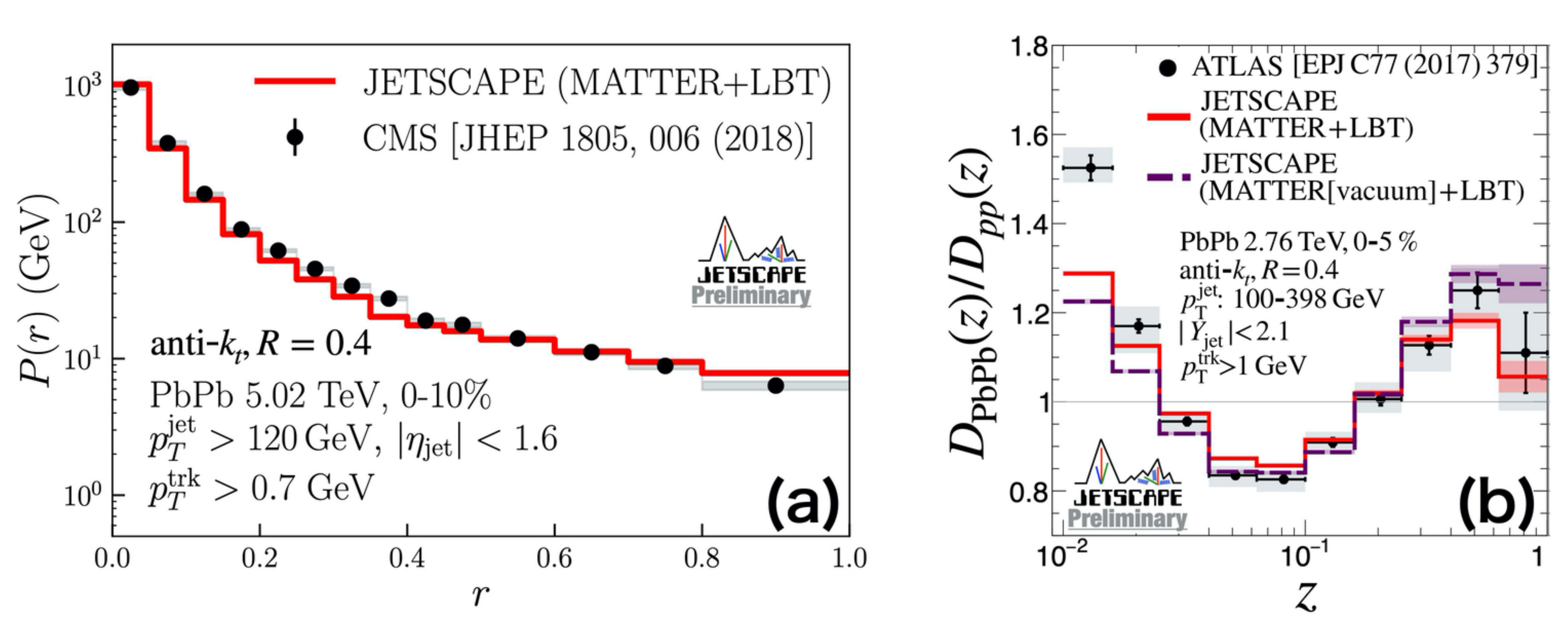}
\end{center}
\vspace{-25pt}
\caption{
(a) Jet shape in PbPb at $5.02~\mathrm{TeV}$ and (b) PbPb/pp jet fragmentation function ratio at $2.76~\mathrm{TeV}$, for MATTER+LBT simulations within JETSCAPE. 
The solid and dashed lines are with and without the medium effects in MATTER. 
The experimental data are from CMS \cite{Sirunyan:2018jqr} and ATLAS \cite{Aaboud:2017bzv}.
}
\label{fig:js_ff}
\end{figure}
Figure~\ref{fig:js_ff} (a) shows the jet shape function for PbPb at $5.02~\mathrm{TeV}$. 
Our result agrees well with the CMS data \cite{Sirunyan:2018jqr}. 
In particular, the broadening at both small and large values of $r$, in which the recoil contribution is significant \cite{Tachibana:2018yae}, is successfully described. 

The modification of the jet fragmentation functions in PbPb at $2.76~\mathrm{TeV}$ is shown in Fig.~\ref{fig:js_ff} (b). 
Our results qualitatively capture the trend of the ATLAS data \cite{Aaboud:2017bzv}: Enhancement at both small $z$ and large $z$. 
We can also see a sizable medium effect in the large-virtuality phase at both small $z$ and large $z$ 
by comparison between the results with and without the medium modification of jets in MATTER. 
The scatterings with the medium in the MATTER phase cause an increase of the splitting rate, especially for hard partons, and reduce the number of large $z$ particles. 
Also, the enhancement at small $z$ is more prominent when the medium effect in MATTER is on. This is because recoils generated early in the jet evolution during the MATTER phase continue to scatter and lose energy through the later LBT stage, causing further enhancement in the small $z$ distribution. 

\subsection{Simulations with the Strongly Coupled Description}
\label{re:strong}
For the simulations with hydrodynamic response in the strongly coupled description, we couple a (3+1)-D viscous hydrodynamic model by MUSIC~\cite{Schenke:2010nt,Schenke:2010rr,Schenke:2011bn} with MATTER+LBT by introducing the source term by the causal formulation. 

\begin{figure}[htbp]
\vspace{-5pt}
\begin{center}
\includegraphics[width=0.998\textwidth,bb=0 0 805 229]{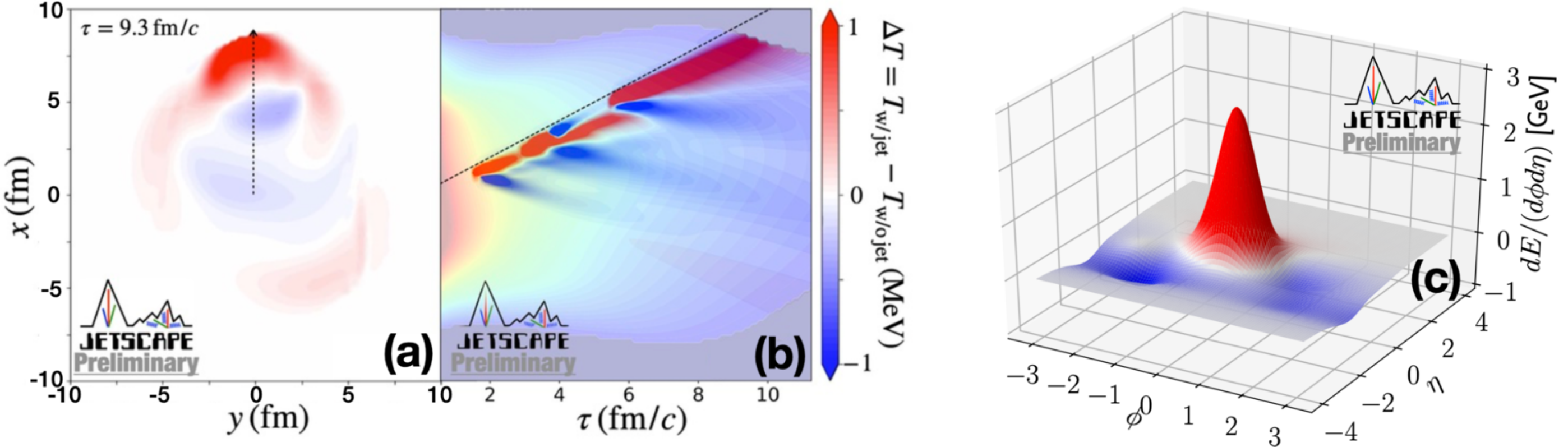}
\end{center}
\vspace{-20pt}
\caption{
Distribution of the fluid temperature difference caused by single-jet propagation in (a) the transverse plane at $(\tau,\eta_{\mathrm{s}})=(9.3~{\mathrm{fm}/c},0)$ and (b) the $\tau$-$x$ plane at $(y, \eta_{\mathrm{s}})=(0,0)$. 
The dashed lines show the path of the jet center. 
(c) Modification of $\phi$-$\eta$ distribution for the energy of hadrons emitted from the medium fluid at the freeze-out. 
Here, the jet direction is set as $(\phi,\eta)=(0,0)$.
}
\label{medmod}
\end{figure}
Figure~\ref{medmod} (a) displays the modification of the medium temperature by jet propagation in the transverse plane. 
One can see a higher temperature region following the jet, which is caused by the positive energy deposition from the soft partons, as well as a lower temperature region caused by a diffusion wake~\cite{Betz:2008ka} which can also be understood in terms of a negative energy deposition from the holes. 
Figure~\ref{medmod} (b) shows the temperature modification in the $\tau$-$x$ plane. 
The causality preservation in our framework can be confirmed by the absence of the modification outside the light-cone.

Finally, we show how the effect of hydrodynamic response appears in the final state in Fig.~\ref{medmod} (c). Jet correlated structures can be seen clearly: Enhancement with a peak around the jet direction and suppression around the peak.

\section{Summary}
\label{summary}
The medium response to jet propagation in the QGP was studied within the JETSCAPE framework. 
Our results based on weak-coupling response describe the jet substructure observables very well. 
However, applying the recoil prescription to the transport process at the typical momentum scale of the thermal medium 
is not consistent with the hydrodynamic description of the bulk medium,  
which has been very successful in explaining a large number of observables in the heavy-ion collisions. 

In our study with hydrodynamic medium response, 
the manifestation of a jet-correlated structure formed in the bulk medium was also shown from a test simulation. 
This clearly demonstrates that the diffusion of energy and momentum from the jet, and the ensuing response of the medium fluid influences jet substuctures. 

To thoroughly understand the contribution of medium response to the jet substructure, 
systematic studies including hydrodynamic response are essential. 
Nevertheless, performing simulations with such fully integrated models is very complex and still challenging. 
Since there is no clear distinction between the jet and the medium background in such models, 
analyses for simulations require 
an elaborated procedure including appropriate background subtraction to construct observables consistently with experimental data. 
In addition, numerical precision greater than that used for the typical modeling of the bulk is needed, as the diffused energy and momentum deposited by jets tends to be tiny compared to the energy and momentum of the bulk medium. 
As a result, accurate simulations for each segment of the model are required for a comprehensive understanding of jet quenching phenomena. 

\vspace{5pt}
\begin{noindent}
\emph{Acknowledgements:} 
These proceedings are supported in part by the National Science Foundation (NSF) within the framework of the JETSCAPE collaboration, under grant numbers ACI-1550300.
\end{noindent}

\bibliographystyle{elsarticle-num}
\biboptions{sort&compress}

\end{document}